\documentclass[reprint,twocolumn]{revtex4}
\usepackage{amsfonts}
\usepackage{amssymb}
\usepackage{amsmath}
\usepackage{color}
\usepackage{hyperref}
\usepackage{latexsym}
\usepackage[dvips]{graphicx}

\usepackage{subfig}
\usepackage{epsf}
\usepackage{multirow}

\newcommand{\half}{\tfrac{1}{2}}
\newcommand{\p}{\partial}

\newcommand{\be}{\begin{equation}}
\newcommand{\ee}{\end{equation}}
\newcommand{\bi}{\begin{itemize}}
\newcommand{\ei}{\end{itemize}}
\newcommand{\bea}{\begin{eqnarray}}
\newcommand{\eea}{\end{eqnarray}}

  \let\eps=\epsilon

\newcommand{\Db}{\bar{D}}

\newcommand{\cD}{\mathcal{D}}

\newcommand{\cR}{\mathcal{R}}

\newcommand{\cZ}{\mathcal{Z}}

\newcommand{\tN}{\tilde{N}}
\newcommand{\ts}{\tilde{\sigma}}

\newcommand{\hN}{\hat{N}}
\newcommand{\hs}{\hat{\sigma}}

\newcommand{\bN}{\bar{N}}
\newcommand{\bs}{\bar{\sigma}}

\begin{document}

%-----------------------------------------------------------------------------
% Title Page
%-----------------------------------------------------------------------------

\preprint{MZ-TH/10-02}

\title{Asymptotically Safe Lorentzian Gravity }% Force line breaks with \\

\author{Elisa Manrique}
\email{manrique@thep.physik.uni-mainz.de}
\author{Stefan Rechenberger}%
  \email{rechenbe@thep.physik.uni-mainz.de}
\author{Frank Saueressig}
 \email{saueressig@thep.physik.uni-mainz.de}
\affiliation{%
 Institute of Physics, Johannes Gutenberg University Mainz,\\
Staudingerweg 7, D-55099 Mainz, Germany
}%

\date{\today}% It is always \today, today,

\begin{abstract}
The gravitational asymptotic safety program strives for a consistent and predictive quantum theory of gravity based on a non-trivial ultraviolet fixed point of the renormalization group (RG) flow. We investigate this scenario by employing a novel functional renormalization group equation which takes the causal structure of space-time into account and connects the RG flows for Euclidean and Lorentzian signature by a Wick-rotation. Within the Einstein-Hilbert approximation, the $\beta$-functions of both signatures exhibit ultraviolet fixed points in agreement with asymptotic safety. Surprisingly, the two fixed points have strikingly similar characteristics, suggesting that Euclidean and Lorentzian quantum gravity belong to the same universality class at high energies.
\end{abstract}

\pacs{04.60.-m,11.10.Hi,11.15.Tk}

\maketitle

% -------------------------------------------------------------------------------------

%-----------------------------------------------------------------------------
%\section{Introduction}
%-----------------------------------------------------------------------------

General relativity provides a reliable and well-tested theory for the gravitational interactions at distances sufficiently large compared to the Planck scale. Many questions concerning fundamental aspects of space, time, and the gravitational interactions are beyond the scope of this classical theory, however, and may only be answered within a quantum theory for gravity. While there are many proposals for such a theory, the final answer is still elusive. This is mainly owed to the fact that quantizing general relativity via standard perturbation theory requires fixing an infinite number of free parameters, indicating that the theory is perturbatively non-renormalizable.

This observation still leaves the possibility that gravity constitutes a renormalizable field theory at the non-perturbative level, a scenario known as asymptotic safety \cite{wein}. The key ingredient in this scenario is a non-Gaussian fixed point (NGFP) of the gravitational renormalization group (RG) flow which controls the behavior of the theory at very high energies and ensures the absence of unphysical UV divergences. Provided that the NGFP comes with a finite number of unstable directions, the resulting fundamental theory is as predictive as a perturbatively renormalizable one \cite{Niedermaier:2006wt}. This non-perturbative renormalizability has already been established for a number of perturbatively non-renormalizable field theories, including gravity in $D=2+\epsilon$ space-time dimensions \cite{Gastmans:1977ad}.

In the more realistic setting of four-dimensional gravity, the main evidence for asymptotic safety originates either from discrete lattice simulations \cite{Ambjorn:2000dv,Ambjorn:2010rx,Hamber:2009mt}, or continuum functional renormalization group methods \cite{Reuter:1996cp,Reuter:2007rv,Percacci:2007sz} which explicitly demonstrated the existence of a suitable NGFP in a variety of approximations \cite{Lauscher:2001ya,FPrefs,Codello:2008vh}.\footnote{A similar interplay between continuum and lattice techniques has already been fruitful in the context of finite-temperature QCD \cite{Braun:2009gm}.} While showing a remarkable agreement, e.g., in the spectral dimension of space-time at long and short distances \cite{Lauscher:2005qz,CDTfrac}, the discrete and continuum approach manifestly differs by a causal structure. In the discrete case causality appears as an essential ingredient for obtaining a suitable classical limit, while the continuum RG computations have all been carried out for generic Euclidean space-times. Thus, understanding of the role of time constitutes an important step in the asymptotic safety program. 

Here, we address this fundamental question utilizing a novel functional renormalization group equation (FRGE) tailored to take the causal structure of space-time into account. The construction relies on the ADM-decomposition of the metric degrees of freedom, which singles out a preferred time-direction and allows to Wick-rotate between Euclidean and Lorentzian signature metrics, facilitating a direct comparison between the two settings. In addition, the underlying foliated structure of space-time makes the new flow equation applicable to Horava-type gravitational theories, where an anisotropy between space and time is essential for the improved renormalization behavior \cite{Horava:2009uw}.

%-----------------------------------------------------------------------------
%\section{The ADM-decomposed flow equation}
%-----------------------------------------------------------------------------
The starting point for imprinting the causal structure is the ADM-decomposition of the $D$-dimensional space-time metric
\be\label{ADMdec}
ds^2 = \epsilon \hN^2 d\tau^2 + \hs_{ij} \left( dx^i + \hN^i d\tau \right) \left( dx^j + \hN^j d\tau \right) \, ,
\ee
which foliates the space-time $M^D = S^1 \times M^d$ into spatial slices $M^d$, coordinatized by $x^i$, $i = 1, \ldots, d = D-1$, and labeled by a preferred ``time''-coordinate $\tau$. 
The $M^d$ carry a positive definite metric $\hat{\sigma}_{ij}(x, \tau)$ and the lapse $\hN$ and shift vector $\hN_i$ connect the coordinate systems on neighboring slices. The signature parameter $\epsilon$ captures space-times with Euclidean $(\epsilon = +1)$ and Lorentzian $(\epsilon = -1)$ signature, respectively. For technical reasons, the $\tau$-direction is taken to be a compact circle with length $T$, which provides an IR cutoff on the quantum fluctuations in this direction.

The investigation of the asymptotic safety scenario requires an analysis of the underlying RG flow, which is conveniently captured by a Wetterich-type FRGE \cite{Wetterich:1992yh,Reuter:1996cp} for the component fields $\{\hN, \hN_i, \hs_{ij}\}$. Its construction starts from the a priori formal functional integral, $\cZ = \int \cD \hN \, \cD \hN_i \, \cD \hs_{ij} \,\exp\left[- S[ \hN, \hN_i, \hs_{ij} ] \right]$, which is assumed to be invariant under $D$-dimensional coordinate transformations Diff$(M^D)$ acting on the component fields through the decomposition \eqref{ADMdec}. The resulting gauge freedom is fixed via the background field method, splitting the quantum field into a fixed background (denoted by a bar) and fluctuations (indicated by a tilde) $\hN = \bN + \tN, \hN_i = \bN_i + \tN_i, \hs_{ij} = \bs_{ij} + \ts_{ij}$. Since the background fields are completely at our disposal, we set $\bN =1, \bN_i = 0$, while keeping $\bs_{ij}(x, \tau)$ unspecified. This choice guarantees the existence of a unique coordinate system in which the quantum fluctuations are confined to the spatial metric $\tilde{\sigma}_{ij}$ \cite{Dasgupta:2001ue}. The gauge-fixing term for this temporal gauge \cite{Teitelboim:1983fk} is 
\be\label{Sgf}
S_{\rm gf} = \tfrac{1}{2} \sqrt{\eps} \int d\tau d^dx \, \sqrt{\bs} \, \left\{ \alpha_L^{-1}\tN^2 +  \alpha_S^{-1} \bs^{ij} \tN_i \tN_j \right\} \,  ,
\ee
where $\alpha_L$ and $\alpha_S $ are gauge-fixing parameters. In Landau-gauge, $\alpha_L , \alpha_S \rightarrow 0$, $S_{\rm gf}$ imposes $\tN=0$ and $\tN_i=0$.  The ghost action resulting from this gauge-choice is found in the standard way and, upon a suitable shift of the ghosts, reads
\be\label{Sghred}
S_{\rm gh} = \sqrt{\eps}\int d\tau d^dx  \, \sqrt{\bs} \,  \left\{ \bar{C} \p_\tau C + \bar{C}_i \p_\tau C^i \right\} \, .
\ee

The final piece in the construction is a scale-dependent IR regulator $\Delta_kS$, which suppresses quantum fluctuations with momenta $p^2 < k^2$ by a $k$-dependent mass term \cite{Reuter:1996cp,Reuter:2007rv}
\be\label{DkS}
\begin{split}
\Delta_kS = & \, \tfrac{\sqrt{\epsilon}}{2}  \int d\tau d^dx \, \sqrt{\bs} \, \\ & \!\!\!\! \!\!\!\!\! \!\!\! \times 
\Big\{ (\tN, \tN_i, \ts_{ij}, \bar{C}, \bar{C}_i) \, \cR_k\, (\tN, \tN_k, \ts_{kl}, C, C^j)^{\rm T}  \Big\} \, .
\end{split}
\ee
The matrix-valued kernel $\cR_k$ vanishes for $k =0$ and is constructed from the background fields only, so that $\Delta_kS$ is quadratic in the fluctuation fields. In contrast to the original construction \cite{Reuter:1996cp}, where the IR-cutoff contains the $D$-dimensional background Laplacian, we require that $\cR_k(\Delta)$ depends on the  \emph{spatial background Laplacian} $\Delta \equiv - \bs^{ij} \Db_i \Db_j$ only. As a consequence, $\cR_k$ acts as an IR-regulator for spatial fluctuations only. This feature is essential, since it avoids the unbounded differential operators, that would occur once  the original construction is carried over to Lorentzian signature. As a consequence, the background gauge symmetry retained by \eqref{DkS} is not Diff($M^D$), but encompasses diffeomorphism invariance on the spatial slices only. 
 
Putting all the pieces together and adding a standard source-term $S_{\rm source}$ for the fluctuation fields, one finally arrives at the $k$-dependent partition function 
\be\label{partfctk}
\begin{split}
\cZ_k = & \int \cD \tN \, \cD \tN_i \, \cD \ts_{ij} \, \cD \bar{C} \, \cD C \, \cD\bar{C}_i \, \cD C^i  \\ & \,  \, \,\, \, \, \times e^{- S\left[ \hN, \hN_i, \hs_{ij} \right] - S_{\rm gf} - S_{\rm gh} - \Delta_kS - S_{\rm source}} \, .
\end{split}
\ee
The effective average action $\Gamma_k[N, N_i, \sigma_{ij}, \bar{\omega}, \omega, \bar{\omega}_i, \omega^i; \bN, \bN_i, \bs_{ij}]$ is constructed as the (modified) Legendre-transform of $\ln \cZ_k$ in the standard way \cite{Wetterich:1992yh,Reuter:1996cp}. It depends on the classical fields $N \equiv \langle \hN \rangle$, $N_i \equiv  \langle \hN_i \rangle$, $\sigma_{ij} \equiv \langle \hs_{ij} \rangle$, $\bar{\omega} \equiv \langle \bar{C} \rangle$, $\omega \equiv \langle C \rangle$, $\bar{\omega}_i \equiv \langle \bar{C}_i \rangle$ and $\omega^i \equiv \langle C^i \rangle$. For completeness, we also define $h_{ij} \equiv \langle \ts_{ij} \rangle$. Taking the $k$-derivative of $\Gamma_k$, the desired FRGE can be cast into the standard form
\be\label{FRGE}
k \p_k \Gamma_k = \half {\rm STr} \left[ \left( \Gamma_k^{(2)} + \cR_k \right)^{-1} k \p_k \cR_k \right] \, .
\ee
Here, $\Gamma_k^{(2)}$ is the second variation of $\Gamma_k$ with respect to the fluctuations and the STr encompasses an integration over spatial loop-momenta together with a trace in field space.
 With the kernel $\cR_k$ chosen properly, eq.\ \eqref{FRGE} encodes the RG flow on the space of all interactions preserving spatial background diffeomorphism invariance. Its solutions interpolate continuously between the bare action for $k \rightarrow \infty$ and the standard effective action $\Gamma = \Gamma_{k=0}[N = \bar{N}, N_i = \bar{N}_i, \sigma_{ij} = \bar{\sigma}_{ij}, \bar{\omega} = \omega = \bar{\omega}_i = \omega^i = 0]$, provided these limits exist.

%-----------------------------------------------------------------------------
%\section{Results}
%-----------------------------------------------------------------------------
We now solve \eqref{FRGE} for a two-dimensional truncation ansatz spanned by the ADM-decomposed Einstein-Hilbert action supplemented by the classical gauge-fixing and ghost terms
\be\label{trunc}
\begin{split}
\Gamma_k = & \tfrac{\sqrt{\epsilon}}{16 \pi G_k}  \! \int \! d\tau d^dx \,  N \sqrt{\sigma} \,  \Big\{ - R + 2 \Lambda_k \\ &
+ \eps^{-1} K_{ij} \left[ \sigma^{ik} \sigma^{jl} - \sigma^{ij} \sigma^{kl} \right] K_{kl} \Big\} + S_{\rm gf} + S_{\rm gh} \, . 
\end{split}
\ee
Here, $K_{ij} = (2 N)^{-1} \left[ \p_\tau \sigma_{ij} - D_i N_j - D_j N_i \right]$ is the extrinsic curvature and $R \equiv {}^{(d)}R(\sigma)$ denotes the curvature scalar on $M^d$. Eq.\ \eqref{trunc} contains a scale-dependent Newtons constant $G_k$ and cosmological constant $\Lambda_k$, but no running couplings that could encode a breaking of Diff($M^D$) in the effective action. Such terms will be investigated elsewhere.

Substituting \eqref{trunc} into \eqref{FRGE}, the $\beta$-functions for the dimensionless couplings $g_k = G_k k^{d-1}$ and $\lambda_k = \Lambda_k k^{-2}$ together with the anomalous dimension of Newtons constant, $\eta_N \equiv G_k^{-1} k\partial_k G_k$, can be read off from the coefficients of the volume and intrinsic curvature terms.  The extraction of these monomials from the operator trace proceeds as follows. First, $\Gamma_k^{(2)}$ is constructed by expanding $\Gamma_k$ around the background $\bN, \bN_i, \bs_{ij}$ to quadratic order in the fluctuations.
Subsequently, we adopt Landau-gauge, which eliminates the fluctuations in the lapse and shift vector. The remaining fluctuation fields are then Fourier-expanded along the circle-direction, which gives rise to infinite sums over Matsubara frequencies. This structure is well known in the context of quantum field theory at finite temperature \cite{Tetradis:1992xd} and, in the gravitational setting, sums the contributions of the Kaluza-Klein states originating from the compactification of the $D$-dimensional theory on a circle. The resulting operator structure of $\Gamma_k^{(2)}$ can be simplified further by choosing  $\bs_{ij}$ as the time-independent metric on the $d$-sphere, which suffices to track the required interaction monomials. Following \cite{Lauscher:2001ya}, the remaining non-minimal operators are eliminated by a transverse-traceless decomposition of the metric fluctuations $h_{ij}(x) \mapsto \left\{ h_{ij}^{\rm T}(x) ,\xi_i^{\rm T}(x), \sigma(x), h(x) \right\}$, where the Jacobians are compensated by a suitable spectral redefinition.  Finally, the IR-regulator  $\cR_k$ is constructed as a Type I cutoff \cite{Codello:2008vh}, implementing the rule $\Delta \mapsto \Delta + R_k(\Delta)$ for the spatial Laplacians. For technical simplicity, we will work with the optimized cutoff \cite{Litim:2001up}, $R_k = k^2 (1 - \Delta/k^2) \theta(1 - \Delta/k^2)$.  With this form of $\cR_k$, the transverse vector $\xi_i^{\rm T}$ and ghost fluctuations do not contribute to the flow. Consequently, the STr splits into a contribution from the transverse-traceless tensor and scalar fluctuations, $\partial_t\Gamma_k = T^\mathrm{TT} + T^0$. The operator traces with respect to $\Delta$ are conveniently evaluated with the early-time expansion of the heat-kernel on $M^d$
\be\label{TTcon}
\begin{split}
T^\mathrm{TT} = & \, \tfrac{\sqrt{\epsilon} \,k^d\, d_{\rm 2T} }{(4 \pi)^{d/2}} \, \sum_n \int d^dx \sqrt{\sigma}  \Big[
  q^{1}_{d/2}(w_{\rm 2T})  \\ & \!\!\!\!\!  
+  \tfrac{R}{k^2} \left( \tfrac{1}{6} q^{1}_{d/2-1}(w_{\rm 2T}) - \tfrac{d^2-3d+4}{d(d-1)} q^{2}_{d/2}(w_{\rm 2T})  \right)
\Big] \, , \\
T^0 = & \, \tfrac{\sqrt{\epsilon} \, k^d}{(4 \pi)^{d/2}}   \, \sum_n \int d^dx \sqrt{\sigma}  \bigg[
q^{1}_{d/2}(w_0) 
\\ & \, + \tfrac{1}{6} \tfrac{R}{k^2} q^{1}_{d/2-1}(w_0)
- \tfrac{d-2}{2 d \lambda_k} \,  \tfrac{R}{k^2} \bigg\{
   q^{1}_{d/2}(w_0) \\ & \, 
- \left( \tfrac{3}{2 \epsilon} m^2 n^2 - \tfrac{4 (d-3)}{d-2} \lambda_k + 1 \right)
\, q^{2}_{d/2}(w_0) 
\bigg\}
\bigg] \, .
\end{split}
\ee
Here, $d_{\rm 2T} = \half (d+1)(d-2)$ and the arguments of the dimensionless threshold functions $q^{p}_{n}(w) \equiv \left( \tfrac{1}{\Gamma(n+1)} - \tfrac{1}{2\Gamma(n+2)} \eta_N \right) (1+w)^{-p} $ are given by
\be\label{warg}
\begin{split}
w_{\rm 2T} = &  \tfrac{1}{2 \epsilon} \, m^2 n^2 - 2 \lambda_k \, , \\
w_0 = & - \tfrac{1}{4 \lambda_k} \left( \tfrac{1}{\epsilon} \,  m^2 n^2  - 4 \lambda_k \right) \left( \tfrac{d-1}{d-2} \tfrac{1}{\epsilon} \, m^2 n^2  - 2 \lambda_k \right) \, . 
\end{split}
\ee
The $\beta$-functions
 $\p_t g_k = \beta_g(g, \lambda; m)$ and $\p_t \lambda_k = \beta_\lambda(g, \lambda; m)$ are read off from the volume and intrinsic curvature terms. Besides being funtions of $g_k, \lambda_k$ they also depend parametrically on the (dimensionless) Kaluza-Klein mass $m = 2 \pi/(T k)$.

Eq.\ \eqref{TTcon} explicitly shows that the fluctuations along the circle are captured by Matsubara sums with mass $m$. The sign of these mass-terms is fixed by the signature of space-time, i.e., $\epsilon$ only appears in the combination $\epsilon^{-1} m^2 n^2$. Notably, all the sums can be carried out analytically, utilizing $\sum_n (n^2 + x^2)^{-1} = \pi (x \tanh(\pi x))^{-1}$ together with suitable parametric derivatives. For $x^2 >0$ this resummation leads to hyperbolic functions. For $x^2 < 0$ the sums can be analytically continued, resulting in trigonometric terms. The analytic structure of the $\beta$-functions depends on $\epsilon$ and $\lambda$ and changes at the lines $\lambda^{(1)} = - \tfrac{1}{4d^2} (d-1)(d-2)$ and $\lambda^{(2)} = 1/2$. For $\lambda < \lambda^{(1)}$ the Euclidean $\beta$-functions are purely hyperbolic while the Lorentzian ones contain only trigonometric terms. For $\lambda > \lambda^{(2)}$ the situation is exactly reversed. In the ``central region'' $\lambda^{(1)} < \lambda < \lambda^{(2)}$, there is always a mixture of both types of terms.

In view of the asymptotic safety program, we now investigate whether the $\beta$-functions derived above possess a NGFP  
 $\{g_*, \lambda_*\}$, satisfying $\beta_g(g_*, \lambda_*; m) = 0$, $\beta_\lambda(g_*, \lambda_*; m) = 0$ with $g_* > 0 $. 
We first turn to the case where   
$T$ is $k$-independent, which implies a trivial running of $m_k$, $\p_t m_k = - m_k$.
 This supplementary equation has a UV-attractive fixed point at $m_* = 0$. Taking the limit $m \rightarrow 0$, all trigonometric terms in the 
  $\beta$-functions diverge, so that 
 the flow is well-defined in the regions $\lambda < \lambda^{(1)}$ for Euclidean and $\lambda > \lambda^{(2)}$ for Lorentzian signature only. 
The only fixed point appearing in this scenario results from the Euclidean $\beta$-functions and is located at $g_* < 0$. Thus it is not suitable 
for rendering the theory asymptotically safe.

The more natural scenario assumes that $m$ is a constant, $k$-independent number. This corresponds to relating the IR cutoff along the time-circle with the spatial cutoff, $T \propto k^{-1}$. In fact,
imposing that the (Euclidean) spatial and time-like fluctuations are cut off at the same momentum scale requires $T = k^{-1}$. Thus $m = 2 \pi$ is distinguished and we adopt this value in the following. Remarkably, this entails that all trigonometric terms remain finite throughout the central region $\lambda^{(1)} < \lambda < \lambda^{(2)}$. 
\begin{figure}[t!]
\centering
  \subfloat[][\small{Euclidean phase portrait}]{\label{Eucl}\includegraphics[width=0.43\textwidth]{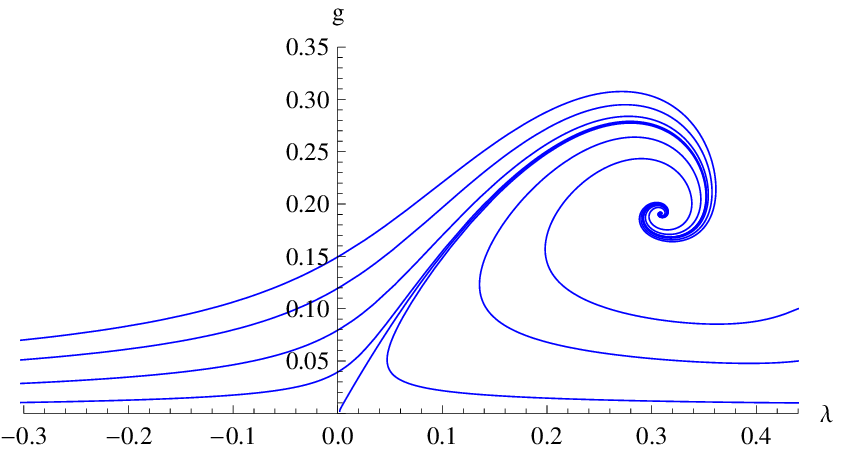}}\\
  \subfloat[][\small{Lorentzian phase portrait}]{\label{Lor}\includegraphics[width=0.43\textwidth]{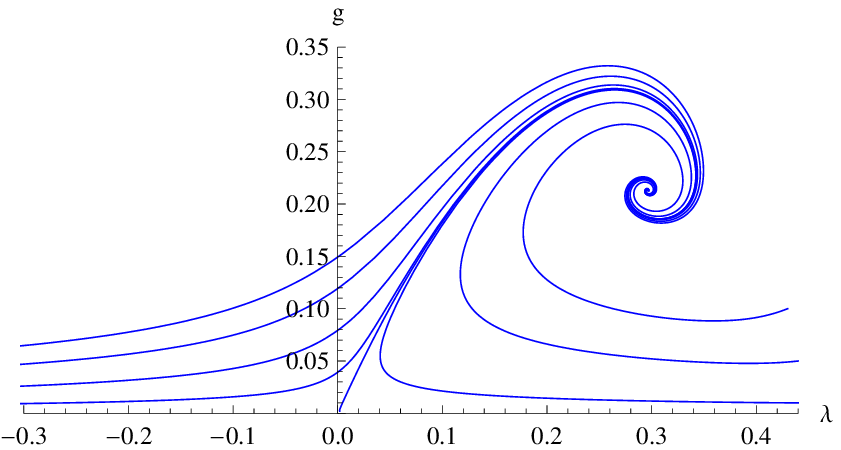}}
 \caption[small]{{Phase portraits obtained from the numerical integration of the Euclidean (left) and Lorentzian (right) $\beta$-functions with $m = 2\pi$. The Euclidean flow diagram completely agrees with the covariant result \cite{Reuter:2001ag}.}} 
\label{phase}
  \end{figure}

Analyzing the fixed point structure of the resulting $\beta$-functions one first encounters the Gaussian fixed point $\{g_*, \lambda_*\} = \{0,0\}$, which corresponds to the free theory and constitutes a saddle-point in the $g$-$\lambda-$plane. Moreover, \emph{both signatures} give rise to a NGFP 
\be\label{FPdata}
\begin{array}{|c||c|c||c|c|}
\eps & g_* & \lambda_* & g_* \lambda_* & \theta_{1,2} \\ \hline
 +1 &  0.19 &  0.31 &  0.059  &  1.07 \pm 3.31 i \\
 -1 &  0.21 &  0.30 &  0.063  &  0.94 \pm 3.10 i \\ \hline 
\end{array}\, ,
\ee
located at $g_* > 0$. The positive real part of the critical exponents $\theta_{1,2}$, defined as the negative eigenvalues of the stability matrix $B_{ij} = \p_j \beta_i$, indicates that the $g$- and $\lambda$-direction both correspond to relevant scaling fields. Thus, \eqref{FPdata} are suitable fixed points for the gravitational asymptotic safety scenario. Notably, the characteristics of the Euclidean NGFP are the same as in the fully covariant computation \cite{Reuter:2007rv}. The most striking feature displayed in \eqref{FPdata} is, however, that the Euclidean and Lorentzian computation gives rise to virtually the same UV-fixed point. This feature is closely related to the analytic structure of the $\beta$-functions in the central region, which contain both hyperbolic and trigonometric terms. This structure persists upon moving from Euclidean to Lorentzian signature, and is likely to be at the heart of this consonance.

As shown by the phase portraits displayed in Fig.\ \ref{phase}, the similarity between the Euclidean and Lorentzian theory continues to hold away from the fixed points: the RG flow of the two theories is virtually indistinguishable. Their sole qualitative difference originates from the trajectories flowing towards $\lambda_k < 0$. 
In the Euclidean case, these can all be continued to the deep infrared $k=0$, while for Lorentzian signature they terminate at a finite value of $k$ when entering the region $\lambda < \lambda^{(1)}$. Again, this difference can be attributed to the analytic structure of the $\beta$-functions. The Euclidean $\beta$-functions are composed out of hyperbolic terms which are well-defined for all values $\lambda < \lambda^{(1)}$. In contrast, the Lorentzian signature gives rise to  trigonometric terms, whose divergences cause the termination of the corresponding RG trajectories.

%-----------------------------------------------------------------------------
%\section{Conclusions}
%-----------------------------------------------------------------------------

In conclusion, the novel causal FRGE presented here provides substantial evidence that the RG flows of \emph{both} Euclidean and Lorentzian gravity feature a non-Gaussian UV fixed point suitable for rendering the theories asymptotically safe. Remarkably, the position and critical exponents of the two fixed points turn out to be virtually identical, indicating that both theories exhibit the same universal UV behavior. The rather surprising result that \emph{signature does not matter} for gravity at very high energies certainly deserves further investigation and may ultimately shed some new light on the fractal properties of space-time at short distances \cite{Lauscher:2005qz,CDTfrac,Carlip:2009kf}. 

\noindent
\emph{Acknowledgments $-$} The research of E.M., S.R.\ and F.S.\ is supported by the Deutsche Forschungsgemeinschaft (DFG)
within the Emmy-Noether program (Grant SA/1975 1-1).

%-----------------------------------------------------------------------------

%-----------------------------------------------------------------------------

\begin{thebibliography}{99}
%-----------------------------------------------------------------------------

\bibitem{wein}
S.~Weinberg 
in \emph{General Relativity, an Einstein Centenary Survey},
S.~W.~Hawking and W.~Israel (Eds.), Cambridge University Press, Cambridge, 1979; 
arXiv:0903.0568 [hep-th]; 
PoS {\bf CD09}, 001 (2009), arXiv:0908.1964 [hep-th].
%

%\cite{Niedermaier:2006wt}
\bibitem{Niedermaier:2006wt}
  M.~Niedermaier, M.~Reuter,
  %``The Asymptotic Safety Scenario in Quantum Gravity,''
  Living Rev.\ Rel.\  {\bf 9}, 5 (2006).



%\cite{Gastmans:1977ad}
\bibitem{Gastmans:1977ad}
  R.~Gastmans, R.~Kallosh, C.~Truffin,
  %``Quantum Gravity Near Two-Dimensions,''
  Nucl.\ Phys.\ B {\bf 133}, 417 (1978);
%\cite{Christensen:1978sc}
%\bibitem{Christensen:1978sc}
  S.~M.~Christensen, M.~J.~Duff,
  %``QUANTUM GRAVITY IN TWO + epsilon DIMENSIONS,''
  Phys.\ Lett.\ B {\bf 79}, 213 (1978).


%\cite{Ambjorn:2000dv}
\bibitem{Ambjorn:2000dv}
  J.~Ambjorn, J.~Jurkiewicz, R.~Loll,
  %``A Nonperturbative Lorentzian path integral for gravity,''
  Phys.\ Rev.\ Lett.\  {\bf 85}, 924 (2000),
  hep-th/0002050.

%\cite{Ambjorn:2010rx}
\bibitem{Ambjorn:2010rx}
  J.~Ambjorn, J.~Jurkiewicz, R.~Loll,
  Annalen Phys.\ {\bf 19}, 186 (2010); arXiv:1004.0352 [hep-th].
  %``Causal Dynamical Triangulations and the Quest for Quantum Gravity,''
  

%\cite{Hamber:2009mt}
\bibitem{Hamber:2009mt}
  H.~W.~Hamber,
  %``Quantum Gravity on the Lattice,''
  Gen.\ Rel.\ Grav.\  {\bf 41}, 817 (2009),
  arXiv:0901.0964 [gr-qc].

%\cite{Reuter:1996cp}
\bibitem{Reuter:1996cp}
  M.~Reuter,
  %``Nonperturbative Evolution Equation for Quantum Gravity,''
  Phys.\ Rev.\  D {\bf 57}, 971 (1998),
  arXiv:hep-th/9605030.
  %%CITATION = PHRVA,D57,971;%%

%\cite{Reuter:2007rv}
\bibitem{Reuter:2007rv}
  M.~Reuter, F.~Saueressig in \emph{Geometric and Topological Methods for Quantum Field Theory},
  H.\ Ocampo, S.\ Paycha and A.\ Vargas (Eds.), Cambridge Univ.\ Press, Cambridge, 2010,
  %``Functional Renormalization Group Equations, Asymptotic Safety, and Quantum, 
  %Einstein Gravity,''
  arXiv:0708.1317 [hep-th].
  %%CITATION = ARXIV:0708.1317;%%

  %\cite{Percacci:2007sz}
\bibitem{Percacci:2007sz}
  R.~Percacci,
  %``Asymptotic Safety,''
  in \emph{Approaches to quantum gravity}, edited by D.\ Oriti, Cambridge Univ.\ Press, Cambridge, 2009, 
  arXiv:0709.3851 [hep-th].



%\cite{Lauscher:2001ya}
\bibitem{Lauscher:2001ya}
  O.~Lauscher, M.~Reuter,
  %``Ultraviolet fixed point and generalized flow equation of quantum gravity,''
  Phys.\ Rev.\ D {\bf 65}, 025013 (2002),
  hep-th/0108040.
%\cite{Lauscher:2001rz}

\bibitem{FPrefs}
  O.~Lauscher, M.~Reuter,
  %``Is quantum Einstein gravity nonperturbatively renormalizable?,''
  Class.\ Quant.\ Grav.\  {\bf 19}, 483 (2002),
  hep-th/0110021;
%\cite{Litim:2003vp}
%\bibitem{Litim:2003vp}
  D.~F.~Litim,
  %``Fixed points of quantum gravity,''
  Phys.\ Rev.\ Lett.\  {\bf 92}, 201301 (2004),
  hep-th/0312114;
%\cite{Codello:2007bd}
%\bibitem{Codello:2007bd}
  A.~Codello, R.~Percacci, C.~Rahmede,
  %``Ultraviolet properties of f(R)-gravity,''
  Int.\ J.\ Mod.\ Phys.\ A {\bf 23}, 143 (2008),
  arXiv:0705.1769 [hep-th];
%\cite{Benedetti:2009rx}
%\bibitem{Benedetti:2009rx}
  D.~Benedetti, P.~F.~Machado, F.~Saueressig,
  %``Asymptotic safety in higher-derivative gravity,''
  Mod.\ Phys.\ Lett.\ A {\bf 24}, 2233 (2009),
  arXiv:0901.2984 [hep-th].

%\cite{Codello:2008vh}
\bibitem{Codello:2008vh}
  A.~Codello, R.~Percacci, C.~Rahmede,
  %``Investigating the Ultraviolet Properties of Gravity with a Wilsonian
  %Renormalization Group Equation,''
  Annals Phys.\  {\bf 324}, 414 (2009),
  arXiv:0805.2909 [hep-th].
  %%CITATION = APNYA,324,414;%%



%\cite{Braun:2009gm}
\bibitem{Braun:2009gm}
  J.~Braun, L.~M.~Haas, F.~Marhauser {\it et al.},
  %``Phase Structure of Two-Flavor QCD at Finite Chemical Potential,''
  Phys.\ Rev.\ Lett.\  {\bf 106}, 022002 (2011),
  arXiv:0908.0008 [hep-ph].
%\cite{Bazavov:2009zn} 

%\cite{Lauscher:2005qz}
\bibitem{Lauscher:2005qz}
  O.~Lauscher, M.~Reuter,
  %``Fractal spacetime structure in asymptotically safe gravity,''
  JHEP {\bf 0510}, 050 (2005),
  hep-th/0508202.

%\cite{Ambjorn:2005qt}
%\cite{Ambjorn:2005db}
\bibitem{CDTfrac}
  J.~Ambjorn, J.~Jurkiewicz, R.~Loll,
  %``Spectral dimension of the universe,''
  Phys.\ Rev.\ Lett.\  {\bf 95}, 171301 (2005),
  hep-th/0505113; Phys.\ Rev.\ D {\bf 72}, 064014 (2005),
  hep-th/0505154.

   
  %\cite{Horava:2009uw}
\bibitem{Horava:2009uw}
  P.~Horava,
  %``Quantum Gravity at a Lifshitz Point,''
  Phys.\ Rev.\ D {\bf 79}, 084008 (2009),
  arXiv:0901.3775 [hep-th].
 
\bibitem{Wetterich:1992yh}
  C.~Wetterich,
  % \emph{ Exact evolution equation for the effective potential},
  Phys.\ Lett.\  B {\bf 301}, 90 (1993).
  %%CITATION = PHLTA,B301,90;%%

 
 %\cite{Dasgupta:2001ue}
\bibitem{Dasgupta:2001ue}
  A.~Dasgupta, R.~Loll,
  %``A proper-time cure for the conformal sickness in quantum gravity,''
  Nucl.\ Phys.\  B {\bf 606}, 357 (2001),
  arXiv:hep-th/0103186.
  %%CITATION = NUPHA,B606,357;%%
  
  
%\cite{Teitelboim:1983fk}
\bibitem{Teitelboim:1983fk}
  C.~Teitelboim,
  %``The Proper Time Gauge In Quantum Theory Of Gravitation,''
  Phys.\ Rev.\  D {\bf 28}, 297 (1983).
  %%CITATION = PHRVA,D28,297;%%



%\cite{Tetradis:1992xd}
\bibitem{Tetradis:1992xd}
  N.~Tetradis, C.~Wetterich,
  %``The high temperature phase transition for phi**4 theories,''
  Nucl.\ Phys.\ B {\bf 398}, 659 (1993);
  D.F.~Litim, J.M.~Pawlowski,
  %``Non-perturbative thermal flows and resummations,''
  JHEP {\bf 0611}, 026 (2006),
  hep-th/0609122.

%\cite{Litim:2001up}
\bibitem{Litim:2001up}
  D.~F.~Litim,
  %``Optimized renormalization group flows,''
  Phys.\ Rev.\ D {\bf 64}, 105007 (2001),
  hep-th/0103195.

%\cite{Reuter:2001ag}
\bibitem{Reuter:2001ag}
  M.~Reuter, F.~Saueressig,
  %``Renormalization group flow of quantum gravity in the Einstein-Hilbert
  %truncation,''
  Phys.\ Rev.\  D {\bf 65}, 065016 (2002),
  arXiv:hep-th/0110054.
  %%CITATION = PHRVA,D65,065016;%%


%\cite{Carlip:2009kf}
\bibitem{Carlip:2009kf}
  S.~Carlip,
  %``Spontaneous Dimensional Reduction in Short-Distance Quantum Gravity?,''
  arXiv:0909.3329 [gr-qc].
  
%-----------------------------------------------------------------------------  
\end{thebibliography}
\end{document}